\documentclass[aps,pre,preprint,nobalancelastpage,nofootinbib]{revtex4-2}
\usepackage{amsmath,amssymb,graphicx,xcolor}

\begin{document}
\title{Trapped Acoustic Energy and Resonances in Spherical Scatterers}
\author{Naruna E. Rodrigues}
\affiliation{Instituto de Física de São Carlos, Universidade de São Paulo, São Carlos 13566-590, Brazil}
\author{José Renato Alcarás}
\affiliation{Faculdade de Filosofia, Ciências e Letras de Ribeirão Preto, Universidade de São Paulo, Ribeirão Preto 14040-900, Brazil}
\author{Gilberto Nakamura}
\affiliation{Instituto de Física de São Carlos, Universidade de São Paulo, São Carlos 13566-590, Brazil}
\author{Odemir M. Bruno}
\affiliation{Instituto de Física de São Carlos, Universidade de São Paulo, São Carlos 13566-590, Brazil}
\author{Alexandre S. Martinez}
\affiliation{Faculdade de Filosofia, Ciências e Letras de Ribeirão Preto, Universidade de São Paulo, Ribeirão Preto 14040-900, Brazil}
\begin{abstract}
The effectiveness of biochemical antivirals are vulnerable to mutations, motivating physical approaches. Recent experiments with ultrasound reveal viral disruption at MHz frequencies, yet the mechanism remains unclear. We model viruses as fluid-like inclusions and analyze internal acoustic fields. Exact solutions reveal that impedance mismatch induces resonances even for subwavelength scatterers. Beyond the monopole, higher-order modes trap significant energy with significant resonance peak broadening even in absence of explicit dissipation mechanisms. These findings suggest internal acoustic resonances as a mechanism for viral destabilization and acoustic metamaterial applications.
\end{abstract}
\maketitle

\section{Introduction}
\label{sec:introduction}

The fight against viral infections has long relied on biomolecular strategies \cite{kumari2022critical}, which typically employ specific binding proteins to trigger immune responses or block viral entry into host cells \cite{gordon2020sars, lamb2024mutation}. Their effectiveness stems from the high specificity of binding to viral proteins \cite{boby2023sars}, but this feature also makes them vulnerable to mutation-induced changes at the binding sites \cite{sanderson2023molnupiravir}.
Physical approaches provide an alternative, since they target generic structural or mechanical properties that are less sensitive to mutations. Ultrasound, for instance, has shown potential to inactivate viruses by exciting resonant modes that compromise viral stability. Finite element simulations indicate that SARS-CoV-2 spike-protein bonds support torsional, bending, and rotational modes and  can resonate in the 1–-20 MHz range \cite{wierzbicki2022}.  In vitro experiments confirmed that SARS-CoV-2 envelopes rupture under ultrasound in the 5–-10 MHz range \cite{Veras2022.11.21.517338}. However, the experiments also demonstrated the rupture of other infectious viruses, including H1N1, reducing the chances of resonances tied to specific viral proteins.

Building on these observations and well-established models \cite{anderson1950,feuillade1999,stanton1988,farra1951,Ivansson5}, we develop a theoretical framework for understanding ultrasound-driven destabilization of viral structures. Instead of modeling molecular bonds directly, we treat viruses as localized fluid inclusions and analyze the resulting acoustic scattering. This approach assumes that materials with negligible shear velocity behave as fluid-like scatterers \cite{doolittle, johnson}. In addition,  we also neglect dissipation in order to highligh the primary effects of acoustic resonances in biological media, following the seminal works of Epstein \& Carhart and Allegra \& Hawley, which have shown that losses primarily broaden the acoustic resonance without suppressing them \cite{EpsteinCarhart1953,AllegraHawley1972}.


In acoustic scattering theory, small particles relative to the incident wavelength can strongly influence the energy transfer, with significant contributions arising from the monopole term. However, higher-order modes can build additional resonances on par with the ones from the monopole, storing higher internal energy storage and, thus, internal pressure or stress. 
These higher‑order modes are the acoustic counterparts of electromagnetic Mie resonances, which arise when an incoming field matches the natural modal frequencies of a sub‑wavelength scatterer~\cite{Arruda:10,Arruda:11,Arruda_2012,PhysRevA.87.043841,Arruda:17,doi:10.1080/17455030.2020.1738590}.  Such resonances concentrate energy and boost scattering cross‑sections far beyond Rayleigh’s long‑wavelength prediction~\cite{chew2019acoustic}, even with $ka\ll1$ where the wavenumber $k = 2\pi/\lambda$ and $a$ is the radius of the scattering center.

Here we argue that ultrasound-induced stresses, usually insufficient to disrupt 100–200 nm particles, are amplified by Mie resonances,  enhancing internal fluid velocities and stress distributions, potentially destabilizing the structural integrity of viral scattering centers. By focusing in the internal energy stored inside the scattering centers, we derive closed‑form expressions for wave‑to‑scatterer energy transfer. Our findings reveal that the resonances are driven by the impedance mismatch between the scatterer and the surrounding fluid, suggesting the internal localization of acoustic waves characterized by a residence time $\tau_{\textrm{Wigner}}$, and the dynamical broadening of resonance peaks.  
We further demonstrate that higher-order modes amplify local energy density and scattering strength beyond a monopole-only picture.
The article is structured as follows. In Sec.\ref{sec:scatp}, we present the scattering of acoustic waves by a small fluid sphere in spherical geometry and derive analytical expressions for the energy transferred in single scattering events. Section\ref{sec:results} reports numerical results for the internal energy distribution in various media for spherical scatterers. Finally, Sec.\ref{sec:conclusions} offers concluding remarks, relating Mie resonances to the phenomenon observed in Ref.~\cite{Veras2022.11.21.517338}. 

\section{Acoustic wave scattering by a spherical particle}
\label{sec:scatp}
\begin{figure}[!htbp]
  \centering
  \includegraphics[width=0.9\linewidth]{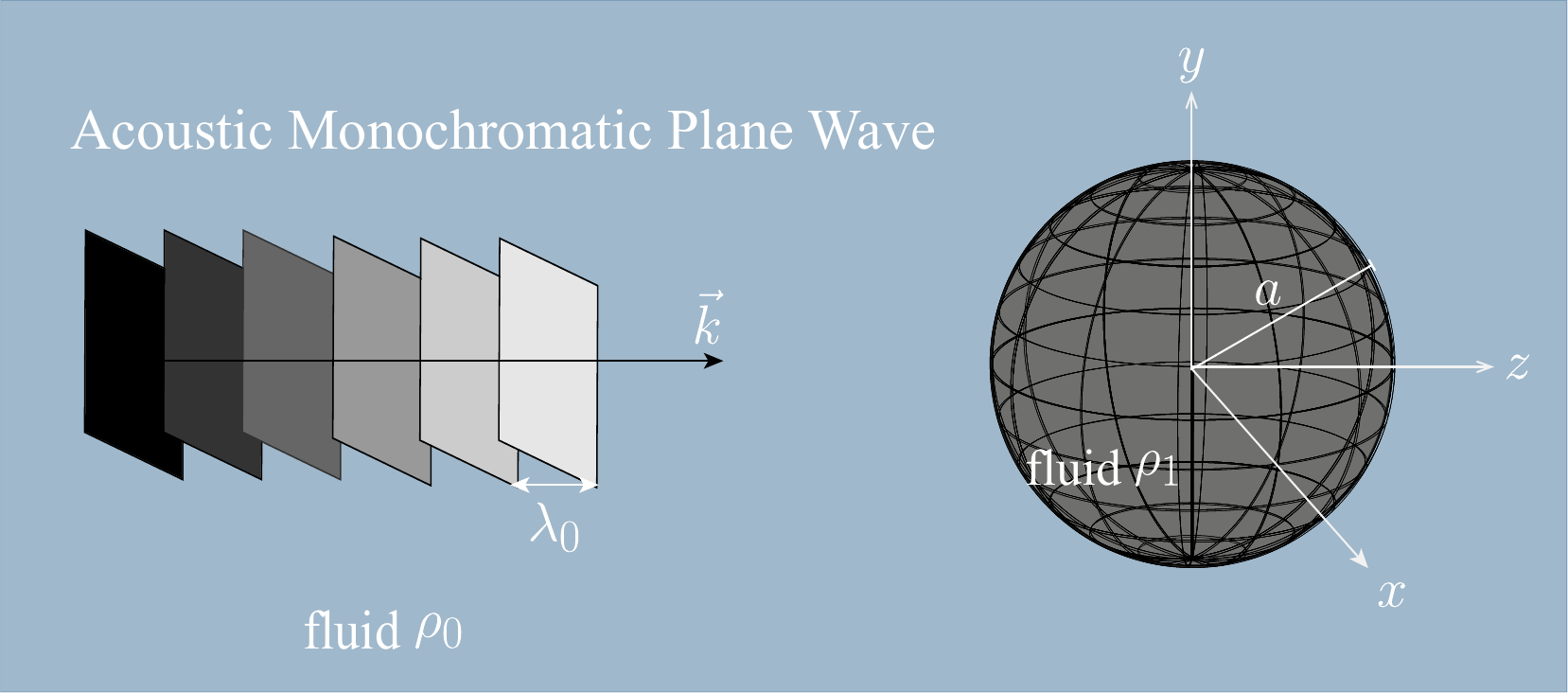}
  \caption{\label{fig:model} Schematic of the scattering setup. A monochromatic acoustic plane wave ($\lambda_{0}$, $\mathbf{k_0}$) propagates along $+z$ strikes a homogeneous sphere (radius $a$, density $\rho{1}$) embedded in a fluid of density $\rho_{0}$.}
\end{figure}

\begin{figure*}[htb]
        \includegraphics[width=0.45\textwidth]{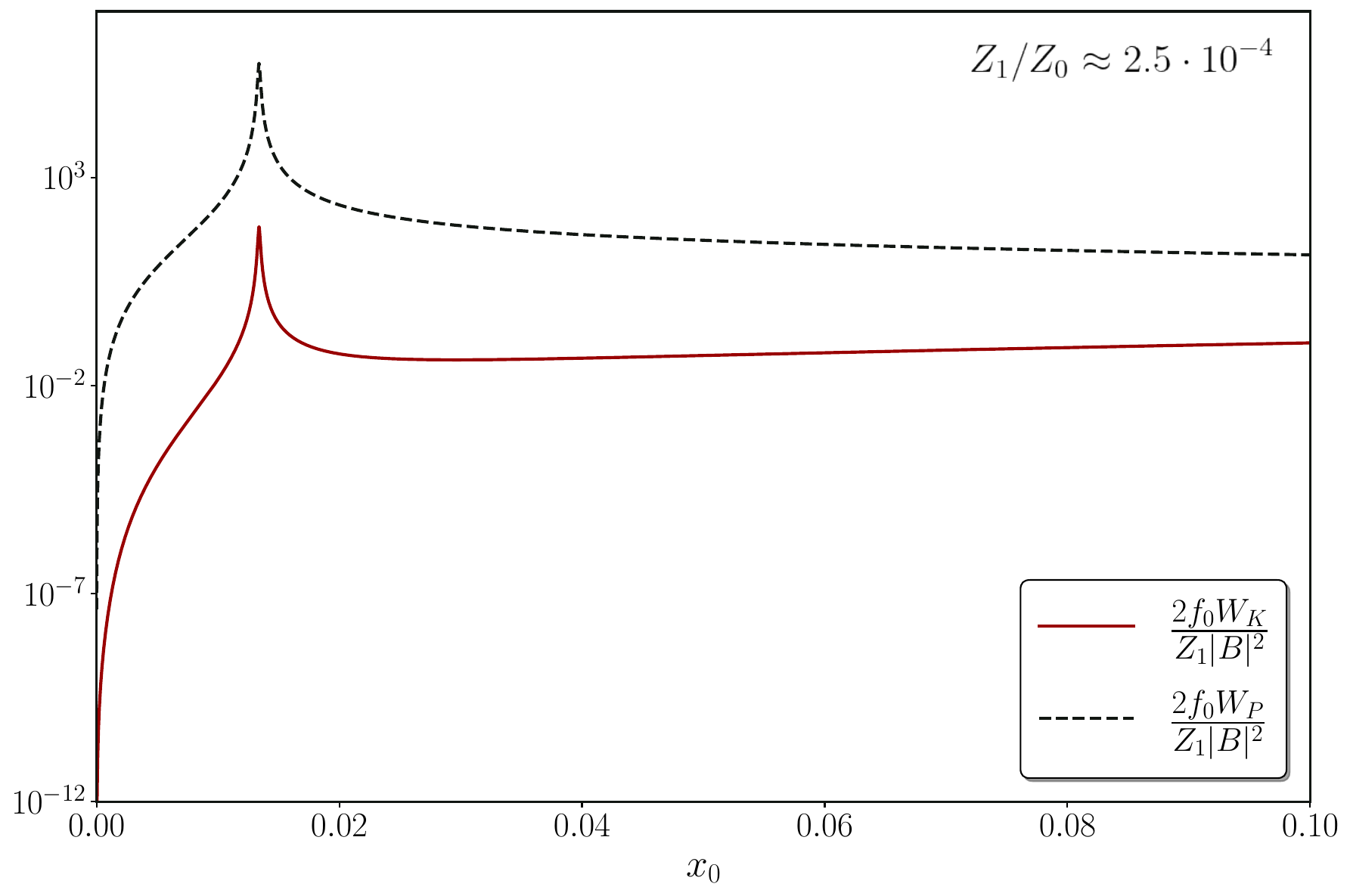}
        \includegraphics[width=0.45\textwidth]{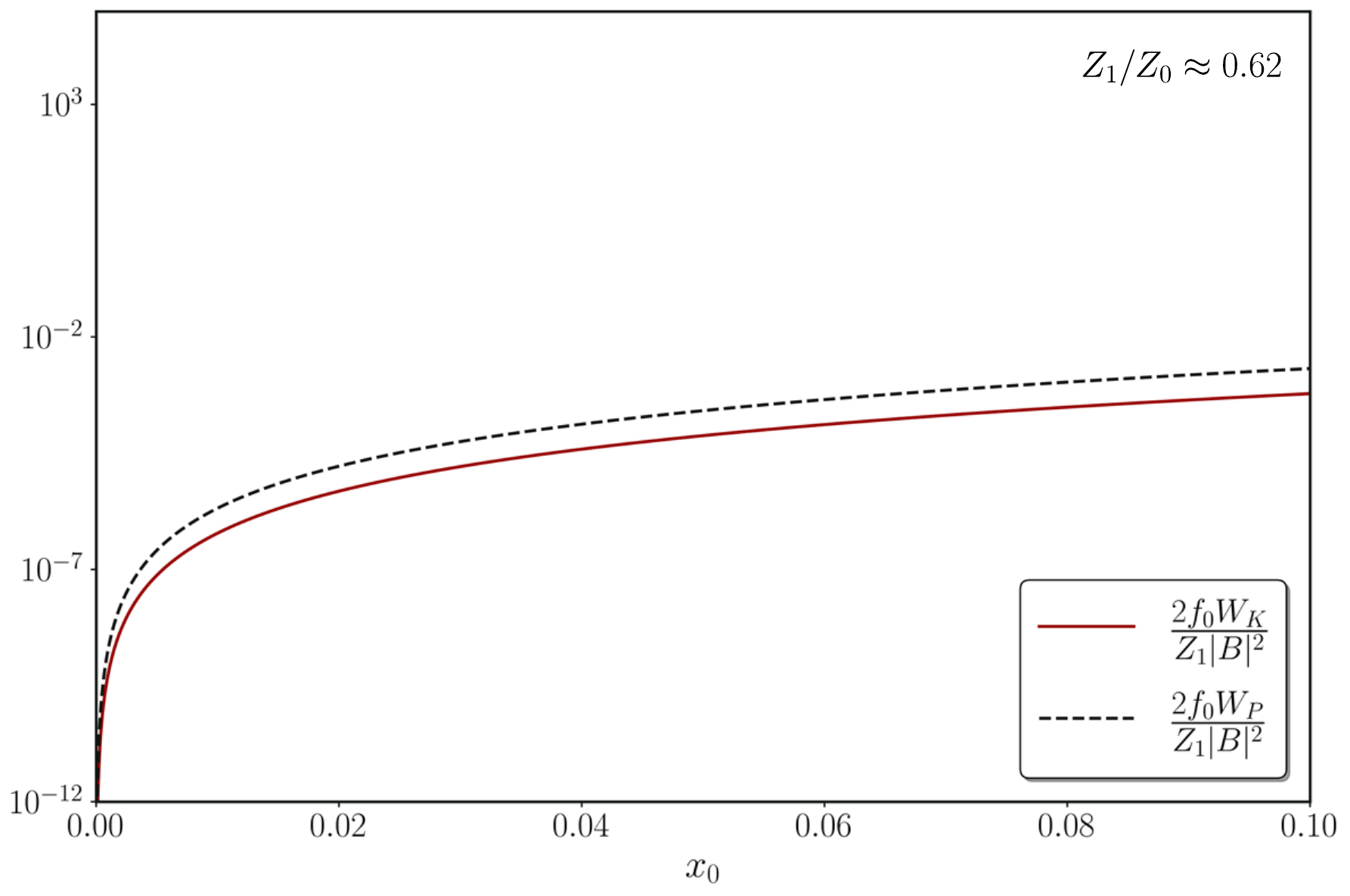}
    \caption{\label{fig:wpwk1-1}Potential and kinetic energies as functions of the dimensionless size parameter $x_0$ for air (left) and petrol (right) in seawater. Small spherical scatterers resonate with incident acoustic waves at low impedance ratio $\left (  Z_1/Z_0 \ll 1\right )$, with potential energy dominating at low $x_0$ by several orders of magnitude. In contrast, the moderate impedance ratio $\left (Z_1/Z_0 \approx 1\right )$ in petrol-seawater fail to support resonances entirely. }
\end{figure*}

The propagation of harmonic acoustic waves in an ideal irrotational fluid with density $\rho_0$ is analyzed~\cite{anderson1950unbound, farra1951}. 
{{ Under these conditions, perturbations in the fluid are described by the scalar potential $\varphi(\mathbf{r})$, which satisfy the Helmholtz equation  $(\nabla^2 + k_0^2) \varphi(\mathbf{r}) = 0$, with wave number $k_0 = \omega/c_0$, angular frequency $\omega$, and sound speed $c_0$  }}~\cite{morse1948,landau1987}.
The pressure field, $p(\mathbf{r}, t) = -\rho \partial_t \left[ e^{-i\omega t} \varphi(\mathbf{r}) \right]$, represents the compression and rarefaction of the fluid as the wave propagates. The velocity field, $\mathbf{v}(\mathbf{r}, t) = e^{-i\omega t} \nabla \varphi(\mathbf{r})$, describes the oscillatory motion of fluid particles caused by the wave.

 In addition, a fluid-like spherical scattering center is positioned at the origin (see Fig.~\ref{fig:model}) with radius $a$, density $\rho_{1}$, sound speed $c_{1}$, and wavenumber $k_{1}=\omega/c_{1}$. The incident acoustic wave, represented by the potential $\varphi_{\text{inc}}(\mathbf{r})$, interacts with the scattering center as it propagates through the surrounding fluid.
Inside the scattering center, the acoustic field is described by a scalar potential $\varphi_1(\mathbf{r})$, while the scattered wave outside is represented by $\varphi_{\text{sc}}(\mathbf{r})$. 
{{All the presented potentials satisfy the Helmholtz equation, with the following boundary conditions:}}
\begin{subequations}
\begin{align}
\rho_0 (\varphi_{\text{inc}} + \varphi_{\text{sc}}) &= \rho_1 \varphi_1,  \label{eq:1a} \\
\frac{\partial}{\partial r} (\varphi_{\text{inc}} + \varphi_{\text{sc}}) &= \frac{\partial \varphi_1}{\partial r}. \label{eq:1b}
\end{align}
\end{subequations}
For a monochromatic incident plane-wave, the solutions take the general form:
\begin{subequations}
    \begin{align}
        \varphi_{\text{inc}}\left(k_0r, \cos{\theta}\right) &= B\sum_{\ell = 0 }^{\infty}\left(2\ell+1\right)i^\ell j_\ell\left(k_0r\right)P_\ell\left(\cos{\theta}\right), \label{eq:incwave}\\
        \varphi_{1}\left(k_1r,\cos{\theta}\right) &= B \sum_{\ell = 0 }^{\infty}b_\ell\left(2\ell+1\right)i^\ell j_\ell\left(k_1r\right)P_\ell \left(\cos{\theta}\right), \label{eq:1wave} \\
        \varphi_{\text{sc}}\left(k_0r,\cos{\theta}\right) &= B \sum_{\ell = 0 }^{\infty}s_\ell\left(2\ell+1\right)i^\ell h_\ell ^{\left(1\right)}\left(k_0r\right)P_\ell\left(\cos{\theta}\right). \label{eq:scwave}
    \end{align}
\end{subequations}
Here B is the wave amplitude, $k_{0,1}$ is the wave number in the surrounding (internal) fluid; $P_\ell(\cos\theta)$ is the Legendre polynomial, $j_{\ell}(z)$ and $h_{\ell}(z)$ are the spherical Bessel and Henkel functions, respectively. 

In terms of the dimensionless parameters $x_0 = k_0 a$ and $x_1 = k_1 a = m x_0$ ($m = c_0/c_1$), together with the boundary conditions, the absorption and scattering coefficients, $b_\ell$ and $s_\ell$, read:
\begin{subequations}
\label{eqs:coeffs}
\begin{align}
b_\ell &= 
-\frac{i x_0^{-2} (\rho_0/ \rho_1)}{
h^{\prime \, (1)}_\ell \left ( x_0 \right )j_\ell\left ( x_1 \right ) - m_t j^{\prime }_\ell\left ( x_1 \right )h_\ell^{\left ( 1 \right )}\left ( x_0 \right )}, \label{eq:blr} \\
s_\ell &= - \frac{j_\ell^{\prime}\left ( x_0 \right )j_\ell\left ( x_1 \right ) - m_t j_\ell\left ( x_0 \right )j_\ell^{\prime}\left ( x_1 \right )}{
h^{\prime \, (1)}_\ell \left ( x_0 \right )j_\ell\left ( x_1 \right ) - m_t j_\ell^{\prime}\left ( x_1 \right )h_\ell^{\left ( 1 \right )}\left ( x_0 \right )}. \label{eq:sl}
\end{align}
\end{subequations}
The impedance ratio $m_t = Z_0/Z_1 =  (\rho_0/\rho_1) m$ governs wave transmission between media, with $m_t \gg 1$ leading to strong relfection ($b_\ell \rightarrow 0$) whereas $m_t \approx 1$ produces negligible scattering ($s_\ell  \rightarrow 0 $). In biological media, where $\rho_1 \approx \rho_0$, transmission depends primarily on the medium organization and the impact on reducing the internal sound speed \cite{cummer2016}. 
Both coefficients share the same denominator in Eq.~(\ref{eqs:coeffs}), which may vanish for special combinations of $m$, $m_t$, and $x_0$. 
These divergences are also observed in far-field problems~\cite{ICESJournal2003,chew2019acoustic}.

The energy of a system composed of an acoustic wave interacting with a scattering center is 
\begin{equation}
W = \int_{V} \frac{|p(\textbf{r}, t)|^2}{2 \rho_1 c_1^2} \, \mathrm{d}V + \int_{V} \frac{\rho_1 |v(\textbf{r}, t)|^2}{2} \, \mathrm{d}V. \label{eq:totalenergy}
\end{equation}
The potential energy arises from pressure variations induced by the sound wave,
while the kinetic energy is associated with the oscillatory motion of fluid particles driven by the passage of the wave.  Using the internal scalar potential $\varphi_1$ from Eq.~\eqref{eq:1wave}, it is possible to explicitly calculate the potential and kinetic energy using Eq.~\eqref{eq:totalenergy} yielding
\begin{widetext}
\begin{subequations}
\begin{align}
    \frac{2f_0 W_P}{Z_1\left|B  \right|^2}  &= x_1^3 \sum_{\ell = 0 }^{\infty} \left | b_\ell \right |^2 \left( 2\ell+1 \right) \left[ j_\ell^2 \left(x_1\right) - j_{\ell-1} \left(x_1\right) j_{\ell+1} \left(x_1\right) \right]. \label{eq:wpfinal},\\
    \frac{2f_0 W_K}{Z_1\left|B  \right|^2}  &= x_1^3 \sum_{\ell = 0 }^{\infty} \left | b_\ell \right |^2 \left( 2\ell+1 \right) \left[ \frac{\ell(2\ell+1)}{x_1^3} I_1^{(\ell)}(x_1) - \frac{2\ell}{x_1^3} I_2^{(\ell)}(x_1) + \frac{1}{2}\left(j_{\ell+1}^2\left(x_1\right) - j_{\ell}\left(x_1\right) j_{\ell+2}\left(x_1\right)\right) \right]. \label{eq:wkfinal} 
\end{align}
\end{subequations}
\end{widetext}
with frequency $f_0$ of the incident wave. The value $(2f_0/Z_1 |B|^2)$ normalizes the expressions for kinetic and potential energy, ensuring they are dimensionless. The integrals $I_1^{(\ell}(x_1) = \int_0^{x_1}j_\ell^2(x) \mathrm{d}x$   and $I_2^{(\ell}(x_1) = \int_0^{x_1} j_\ell(x) j_{\ell+1}(x)\, x \,\mathrm{d}x$ lack an explicit analytical expression and thus must be calculated numerically.
We truncate the summation of the partial-wave series at
\begin{equation}
    \ell_{\text{max}} = 3 + \left\lfloor x_0^{\text{(max)}} + 4.05 \left( x_0^{\text{(max)}} \right)^{1/3}\right\rfloor, \label{eq:lmax}
\end{equation}
following the numerical analysis in Ref.~\cite{kargl1991ray} to ensure the convergence in scattering calculations. The size parameter $x_1 = k_1a$ dictates the number of partial waves required for accurate results. This formula accounts for higher-order contributions, such as leaky Lamb waves, and ensures both accuracy and efficiency in the series summation, particularly in cases involving complex resonances. In this analysis, $x_0^{(\text{max})} = 10$ was the largest value considered, providing a reference for the upper bound on the size parameter for the calculations.

\section{results}
\label{sec:results}
\subsection{Systems with a impedance ratio $Z_1/Z_0\ll 1$}

Acoustic impedance characterizes a medium’s resistance to the transmission of sound waves. A large impedance mismatch between two media results in a strongly reflective interface, thereby impeding the transfer of acoustic energy. This mechanism affects not only the incident wave but also waves generated within a finite scattering center and propagating toward the surrounding fluid. In the absence of dissipation, such waves can become effectively trapped inside the scatterer. Resonances arise when the rate of wave accumulation within the scattering center exceeds the rate in which energy is emitted into the surrounding medium \cite{Arruda:17}.

The large difference in acoustic impedance between medium 0 and medium 1 significantly affects the energy distribution inside the scatterer. 
Here, we study the potential and kinetic energies within the scattering center, using Eqs.~\eqref{eq:wkfinal} and \eqref{eq:wpfinal}, the spherical scattering center was represented by an air bubble - medium 1 (density $\rho_1 = 1.205 \, \text{kg/m}^3 $ and speed of sound $ c_1 = 343 \, \text{m/s} $) - immersed in seawater - medium 0 (the density $ \rho_0 = 1024 \, \text{kg/m}^3 $ and speed of sound $ c_0 = 1522 \, \text{m/s}$. 
The interaction between acoustic waves and the penetrable scatterer is examined as a function of the dimensionless size parameter $x_0$.

\begin{figure}[!htb] 
        \includegraphics[width=0.49\textwidth]{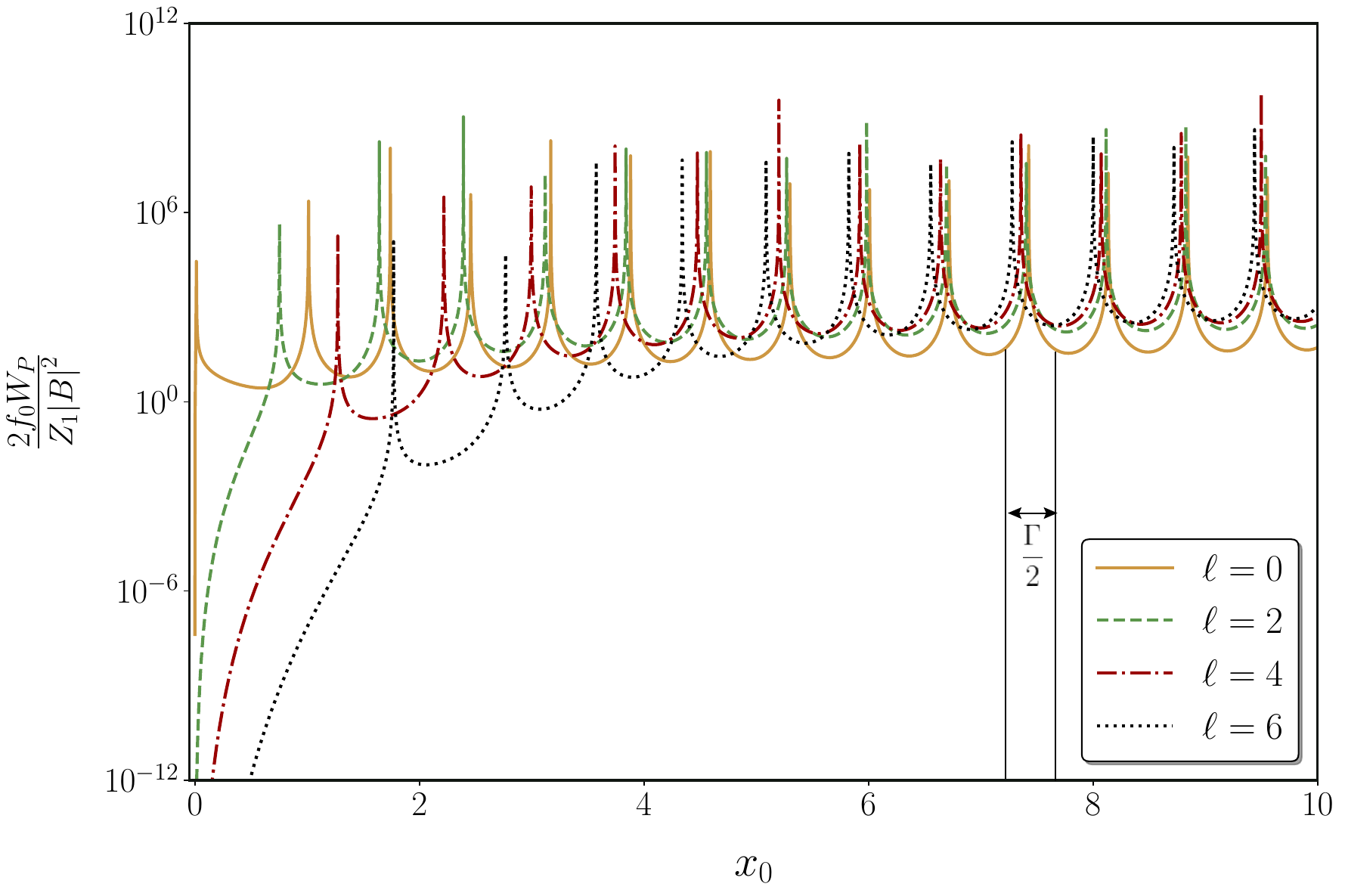}
        \includegraphics[width=0.49\textwidth]{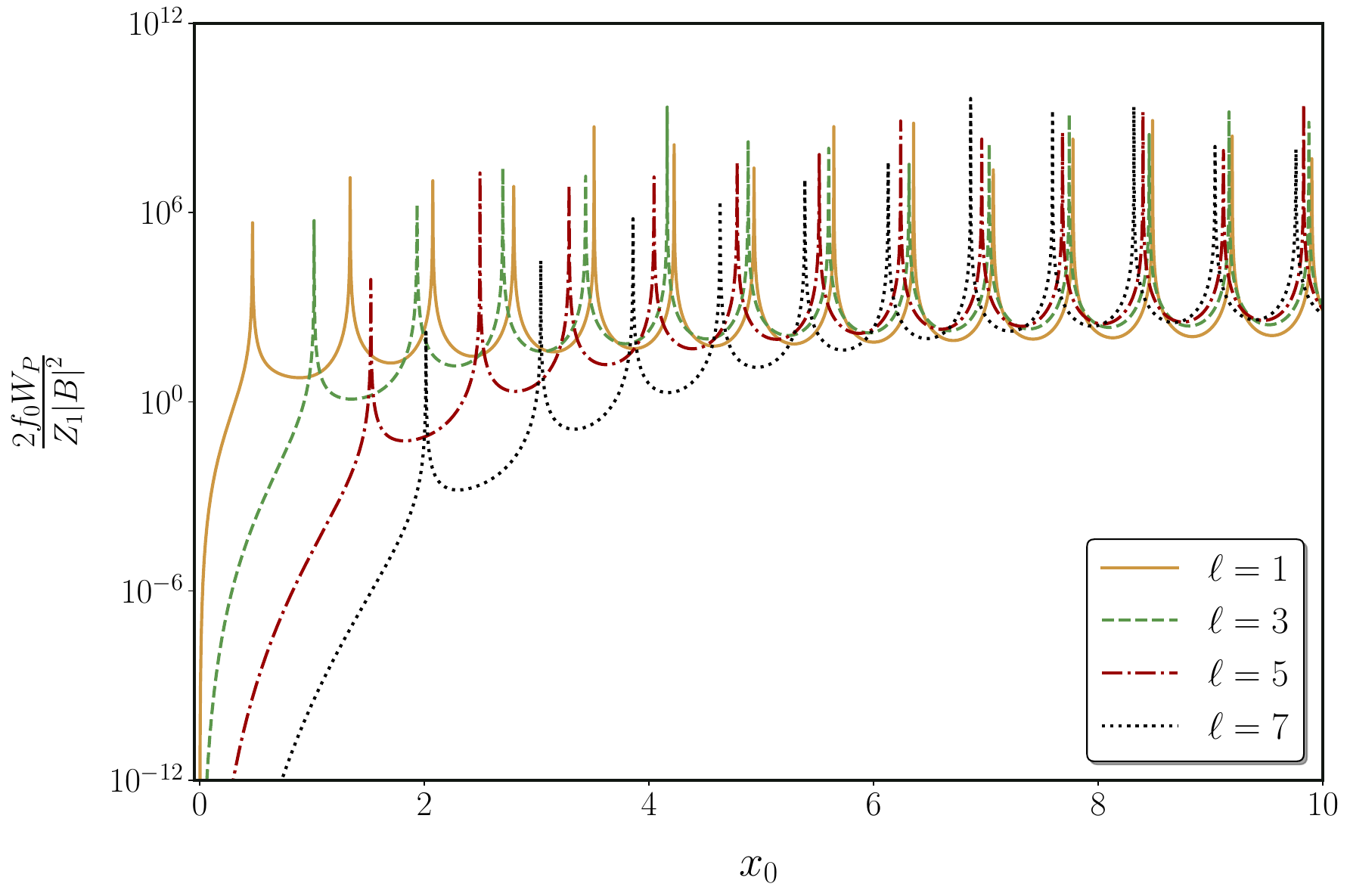}
    \caption{\label{fig:potential_energy1} Potential energy distribution, as a function of the dimensionless size parameter $x_0$, for an air bubble immersed in seawater. Medium 0 corresponds to the surrounding seawater ($\rho_0 = 1024  \text{kg/m}^3$, $c_0 = 1522 \ \text{m/s}$), while medium 1 represents the air bubble ($\rho_1 = 1.205  \text{kg/m}^3$, $c_1 = 343\,  \text{m/s}$). Separate plots are shown for even (top) and odd (bottom) values of $\ell$. The resonance peaks, which grow in number as $x_0$ increases, indicate critical points where the acoustic wave is trapped and energy is deposited within the scattering center.}
\end{figure}

\begin{figure}[htb]
    \centering
        \includegraphics[width=0.49\textwidth]{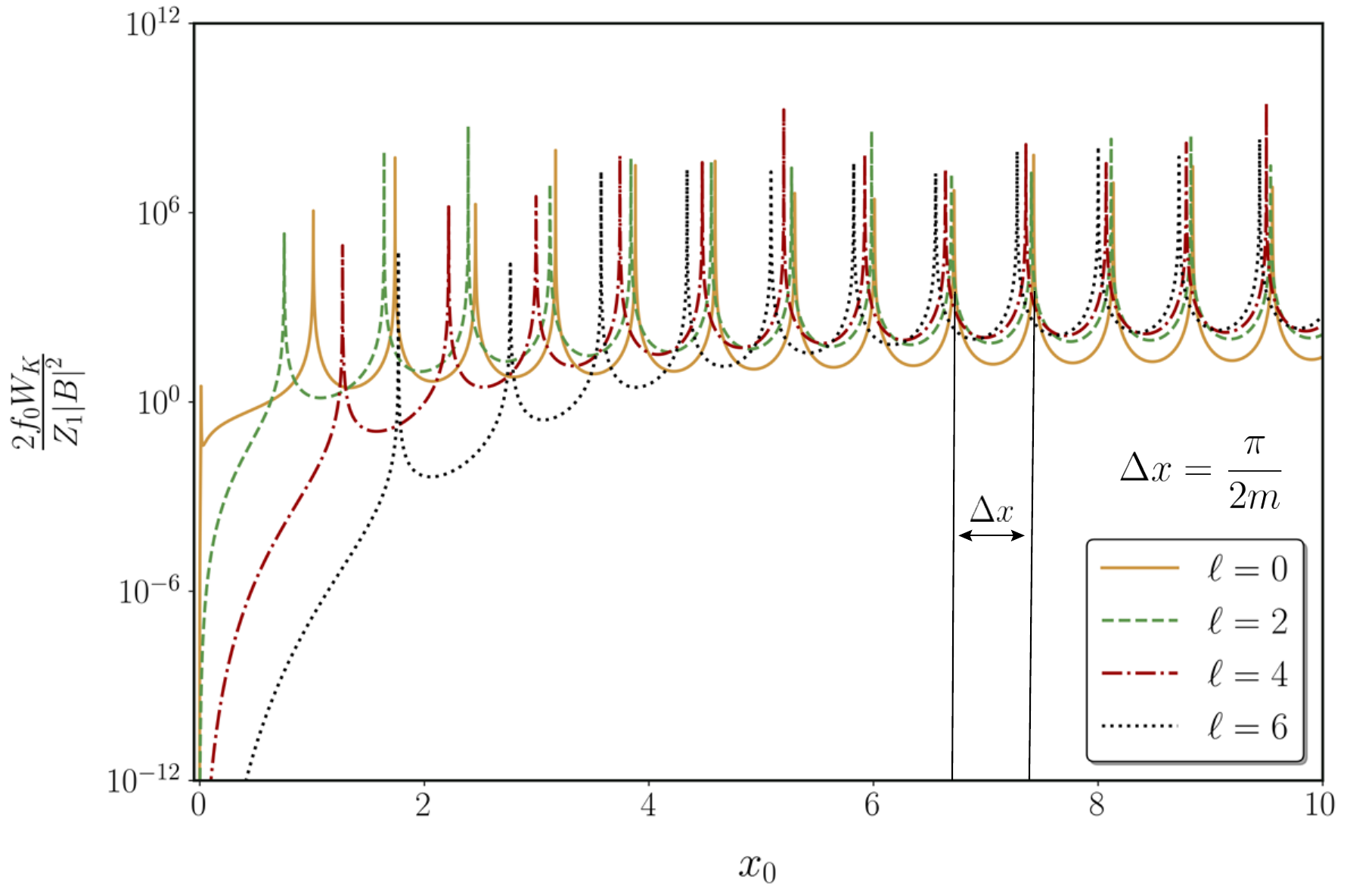}
        \includegraphics[width=0.49\textwidth]{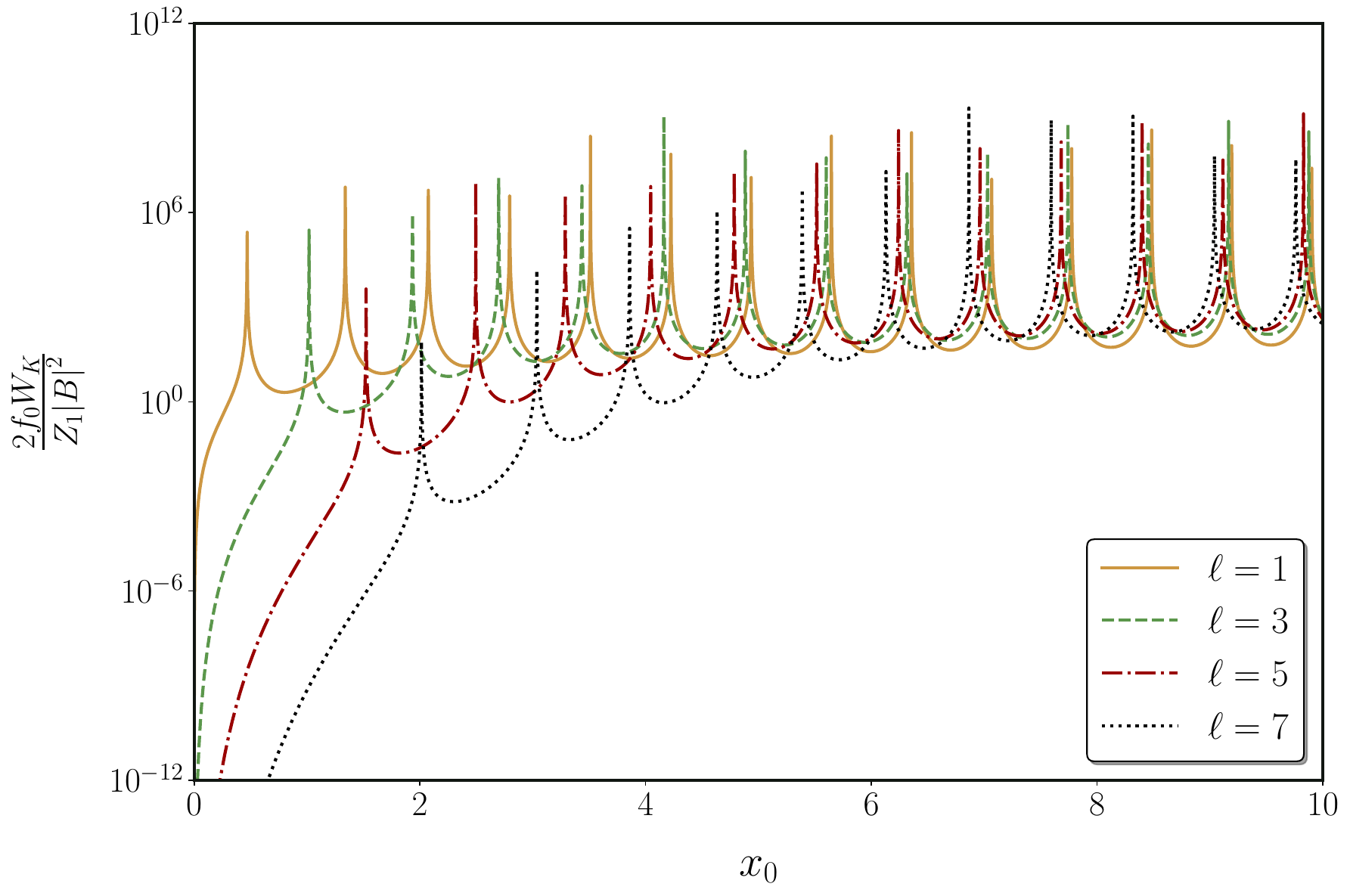}
    \caption{\label{fig:kinetic_energy1} The kinetic energy distribution  as a function of the dimensionless size parameter $x_0$ for the air bubble in seawater. As with the potential energy, the resonance peaks reflect regions of high energy concentration within the scatterer.}
\end{figure}

Fig.~\ref{fig:wpwk1-1} compares the potential and kinetic energies as functions of the dimensionless size parameter $x_0$ for an air bubble immersed (left) and a petrol bubble (right) in seawater. Fig.~\ref{fig:wpwk1-1} (left) shows the energy behavior for $\ell = 0$, highlighting key resonance effects. Resonance occurs when an acoustic wave interacts with a scatterer at specific frequencies, causing constructive interference and a significant amplification of energy inside or around the scatterer. These resonance peaks represent points where the acoustic wave strongly interacts with the scatterer, leading to substantial energy deposition. Notably, a resonance peak appears at $x_0 \ll 1$, corresponding to the monopole mode. This peak arises even though the scatterer is much smaller than the wavelength of the incident wave, primarily due to the impedance contrast between the air bubble and the surrounding seawater. The resonances vanish for impedance ratio close to unity, as show in  Fig.~\ref{fig:wpwk1-1} (right).

The first monopole resonance, shown in Fig.~\ref{fig:wpwk1-1}, at $x_0 = \sqrt{3/mm_t} \approx 0.014$, satisfies
\begin{subequations} \begin{align} \frac{2f_0}{Z_1 |B|^2} \left [ W_K  \right ]_{\text{Small}} &= \frac{m^5 x_0^5}{45} \left[ \frac{m_t}{m} \frac{3}{(3 - m_t m x_0^2)} \right]^2, \\ \frac{2f_0}{Z_1 |B|^2} \left [ W_P  \right ]_{\text{Small}} &= \frac{4m^3 x_0^3}{3} \left[ \frac{m_t}{m} \frac{3}{(3 - m_t m x_0^2)} \right]^2. \end{align}\label{eq:small} \end{subequations}
{{Similarly, the potential energy within the scatterer, as shown in Fig.~\ref{fig:potential_energy1},
exhibits sharp resonance peaks as a function of $x_0$. For larger values of $x_0$, higher-order modes resonate, creating a comb-like structure. As $x_0$ increases further, the frequency of these peaks rises,  as all modes interfere constructively when the scatterer approaches the wavelength size.}}

Likewise, the kinetic energy follows a similar trend, as shown in Fig.~\ref{fig:kinetic_energy1}. Distinct resonance peaks appear as $x_0$ increases, with small-scale resonance clearly visible in the kinetic energy distribution, reflecting efficient energy transfer even for a small scatterer. As $x_0$ increases, the kinetic energy peaks across different values of $\ell$ become far more synchronized and broader, with a characteristic linewidth $\Gamma$.

Resonances are classically described by $(E-E_0-i\Gamma/2)^{-1}$ with $\Gamma$ being associated as a damping mechanism. However, our model has no internal losses: this apparent damping is purely dynamical. Each cavity mode couples to the continuum of outgoing waves, so the surrounding fluid behaves like an absorbing wall at infinity. In Fano’s picture, a trapped mode interferes with a nonresonant background, turning ideal $\delta$ peaks into finite and often asymmetric profiles \cite{PhysRevA.87.043841,Arruda:17, Miroshnichenko2010RMP, CundiffYe2003RMP}. Energy is not dissipated inside the inclusion; it is radiated after an average time $\tau_{\text{Wigner}}$.

In the diffraction limit ($ka \gg 1$), the solutions (\ref{eq:blr}) and (\ref{eq:sl}) simplify, yielding the $\mu$-th resonance position $\mathrm{X}_\ell^\mu = ({\pi}/{2m})\left ( 2\mu+1+\ell \right )$.
From this expression, the spacing between consecutive resonances is $\Delta_{\mu}X = \pi/m$ for fixed $\ell$, and $\Delta_{\ell}X = \pi/(2m)$ for fixed $\mu$. Since $\Gamma \sim 1/\tau_{\text{Wigner}}$ \cite{PhysRevA.87.043841}, it follows that $\tau_{\textrm{Wigner}} / \tau_{\textrm{min}} \sim (m/\pi)$, indicating that the wave remains trapped longer when $m \gg 1$ and minimum cycling time $\tau_{\textrm{min}} = 2\pi a /c_1$. This regime is relevant for certain viral particles \cite{Kol2007}, where the contrast parameter can reach $m_{\textrm{HIV}} \approx o(10^2)$.

\begin{figure}[htb]
\includegraphics[width=0.495\textwidth]{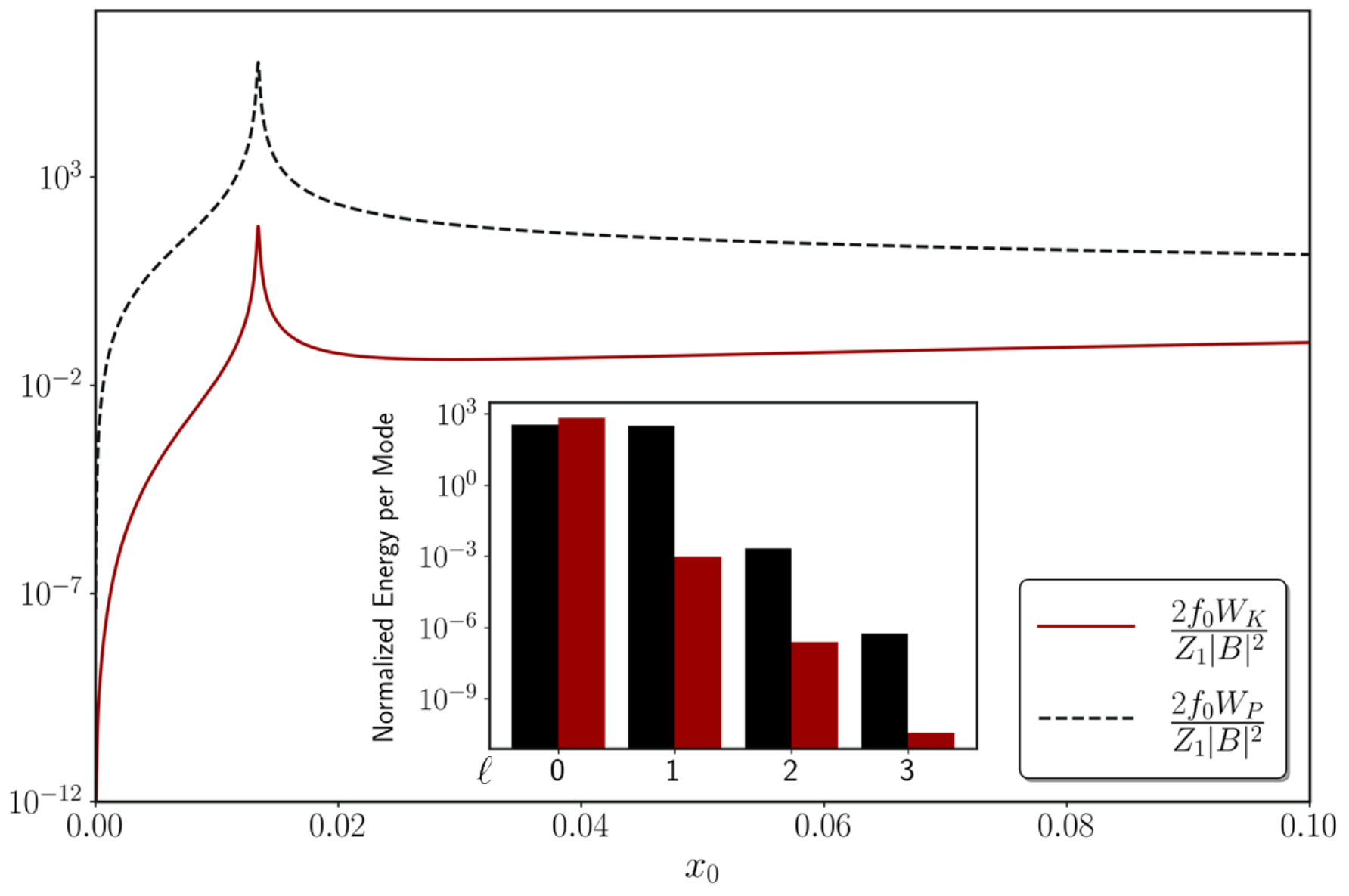}
\includegraphics[width=0.495\textwidth]{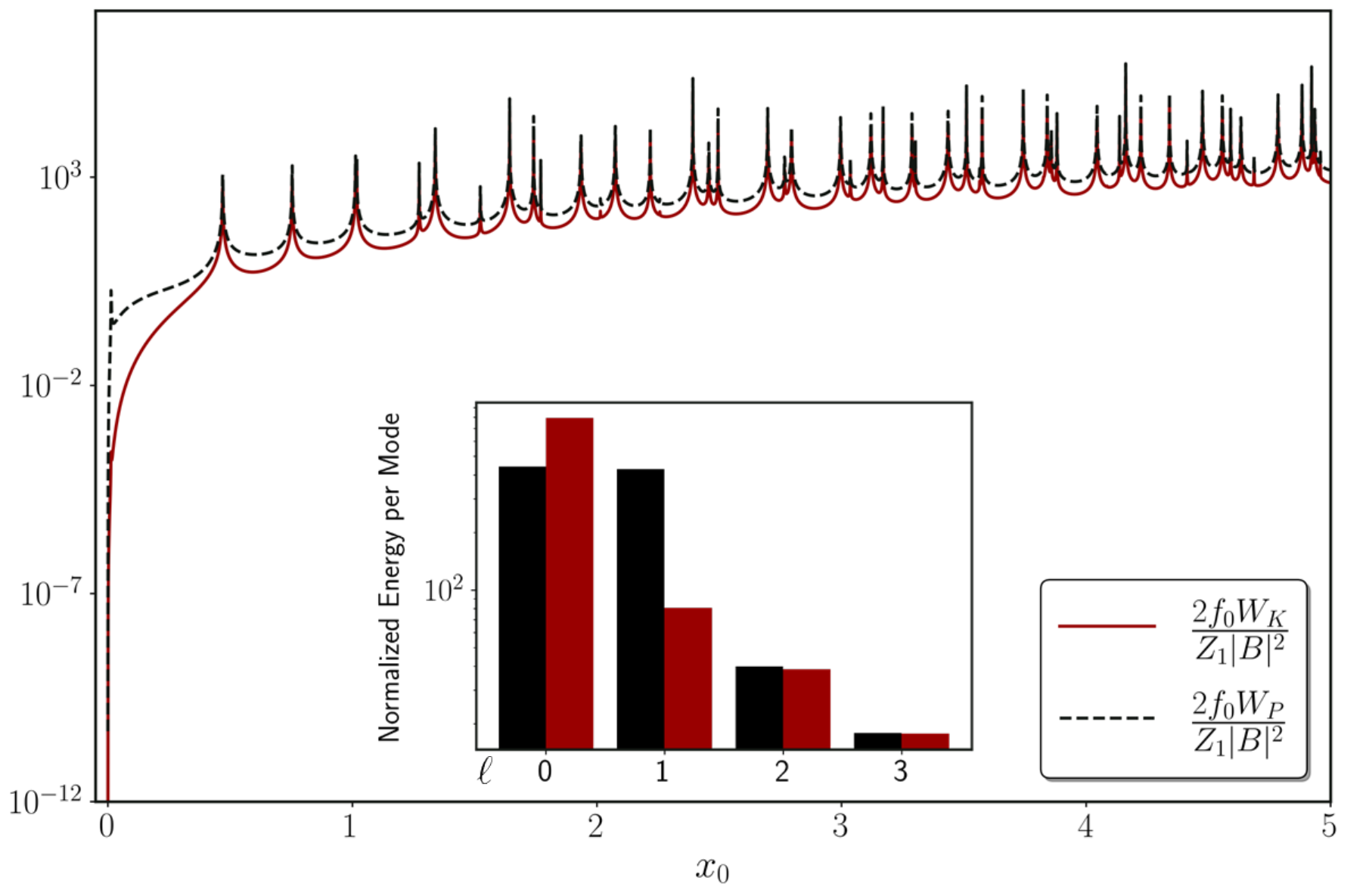}
\caption{\label{fig:energy-contribution} Relative modal contribution. 
Total potentail (black, dashed) and kinetic (red, full) energy in the small (right) and large (left) particle limit, highlighting the resonances.  (inset) Although the monopole remains as the most significant mode overall, the energy histogram reveals a significant contribution from other modes. }
\end{figure}

Figure~\ref{fig:energy-contribution} plots the stored-energy contribution for all multipoles. Both kinetic and potential energy show that higher-order modes ($\ell>0$) have significant contributions even around resonances. Thus, a monopole-only approximation is asymptotically exact when $x_0 \ll 1$; once $x_0 \gg 0$, omitting $\ell>0$ underestimates the stored energy.

The results highlight pronounced resonance behavior. Both energies show sharp peaks for some values of $x_{0}$, including for $x_{0}\ll1$. The strong impedance mismatch between the medium and the scattering center leads to substantial internal energy storage. As $x_{0}$ increases, resonances become denser and other modes contributions cannot be neglacted.

\subsection{Systems with a impedance ratio $Z_1/Z_0 \approx 1$}

In this section, we investigate the behavior of both potential and kinetic energies for a system comprising a petrol bubble submerged in seawater. The petrol bubble, with an acoustic impedance of approximately $968\,750 \, \text{kg/m}^2 \cdot \text{s}$, is immersed in seawater, whose impedance is around $1\,558\,528 \, \text{kg/m}^2 \cdot \,\text{s}$. 
This results in a moderate impedance ratio ($Z_1 / Z_0 \approx 0.62$). 

\begin{figure}[htb]
        \includegraphics[width=0.45\textwidth]{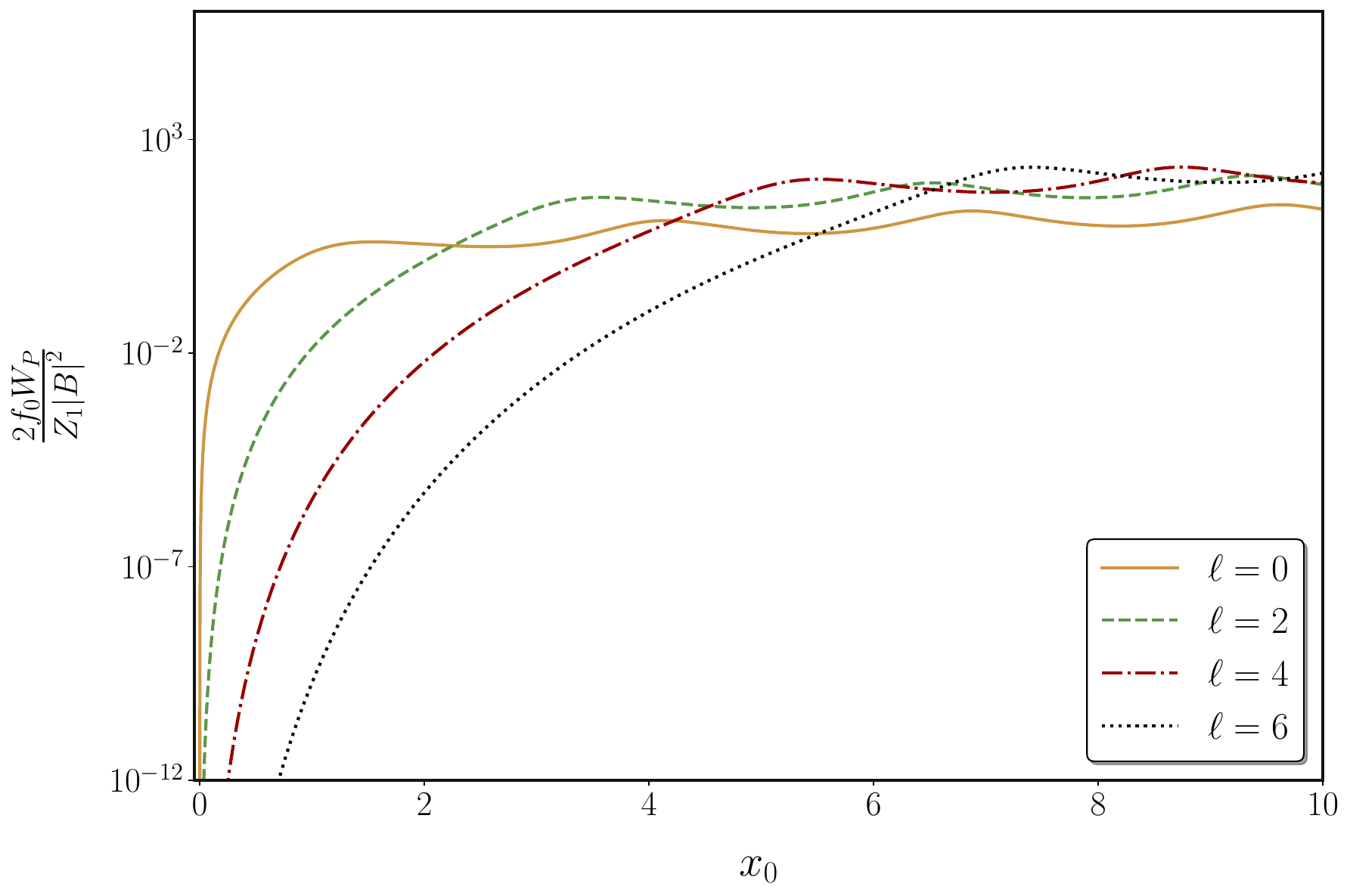}
        \includegraphics[width=0.45\textwidth]{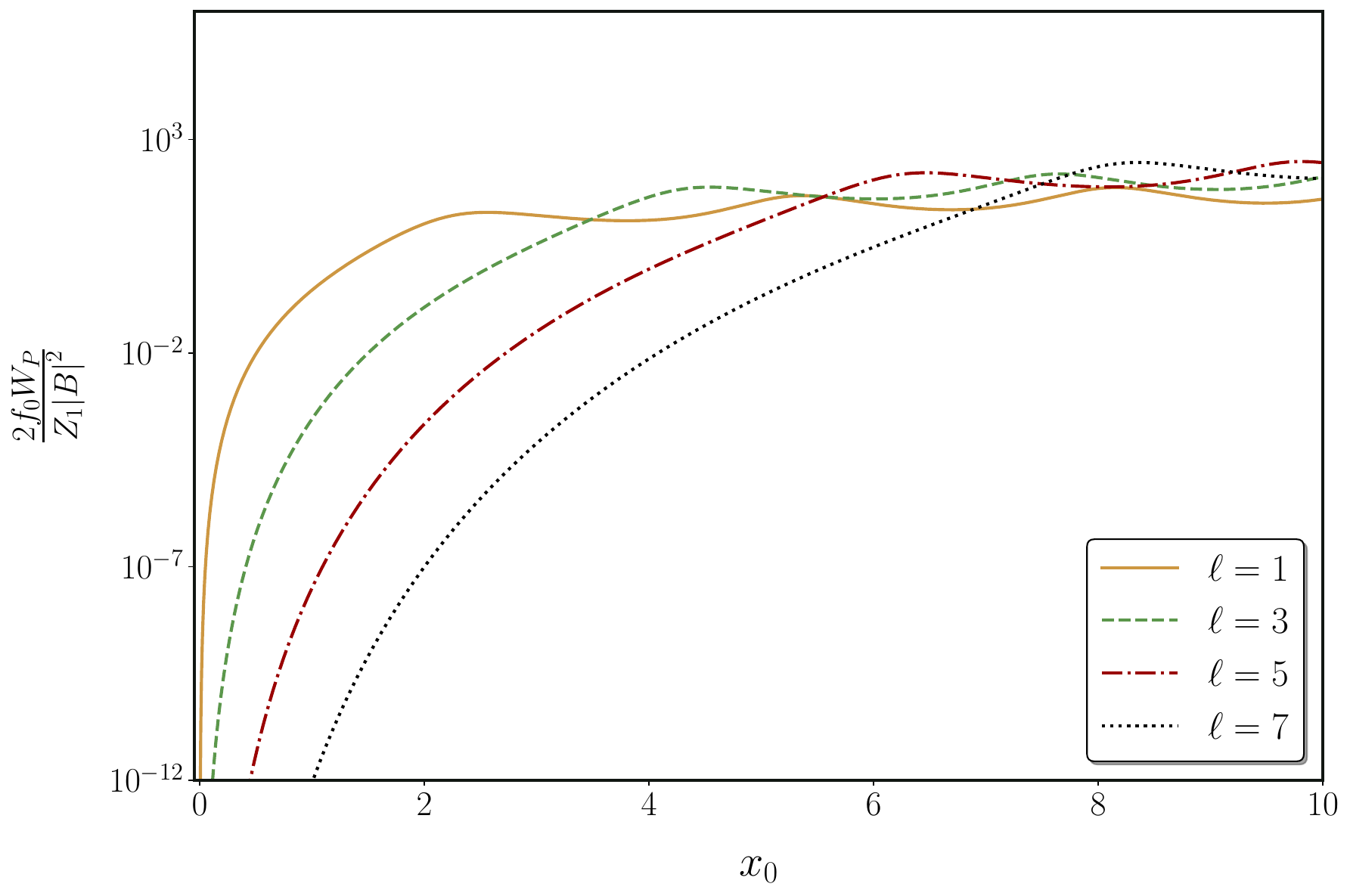}
    \caption{\label{fig:wp2} Potential energy as a function of $x_0$ is given by Eq.~\eqref{eq:wpfinal}, plotted for even ($\ell = 0, 2, 4, 6$) and odd ($\ell = 1, 3, 5, 7$) modes. The system consists of seawater (medium 0: $\rho_0 = 1024\, \text{kg/m}^3$, $c_0 = 1522\, \text{m/s}$) and a petrol bubble (medium 1: $\rho_1 = 968.75\,\text{kg/m}^3$, $c_1 = 343\, \text{m/s}$). No significant resonance peaks are observed.}
\end{figure}

\begin{figure}[htb]
        \includegraphics[width=0.45\textwidth]{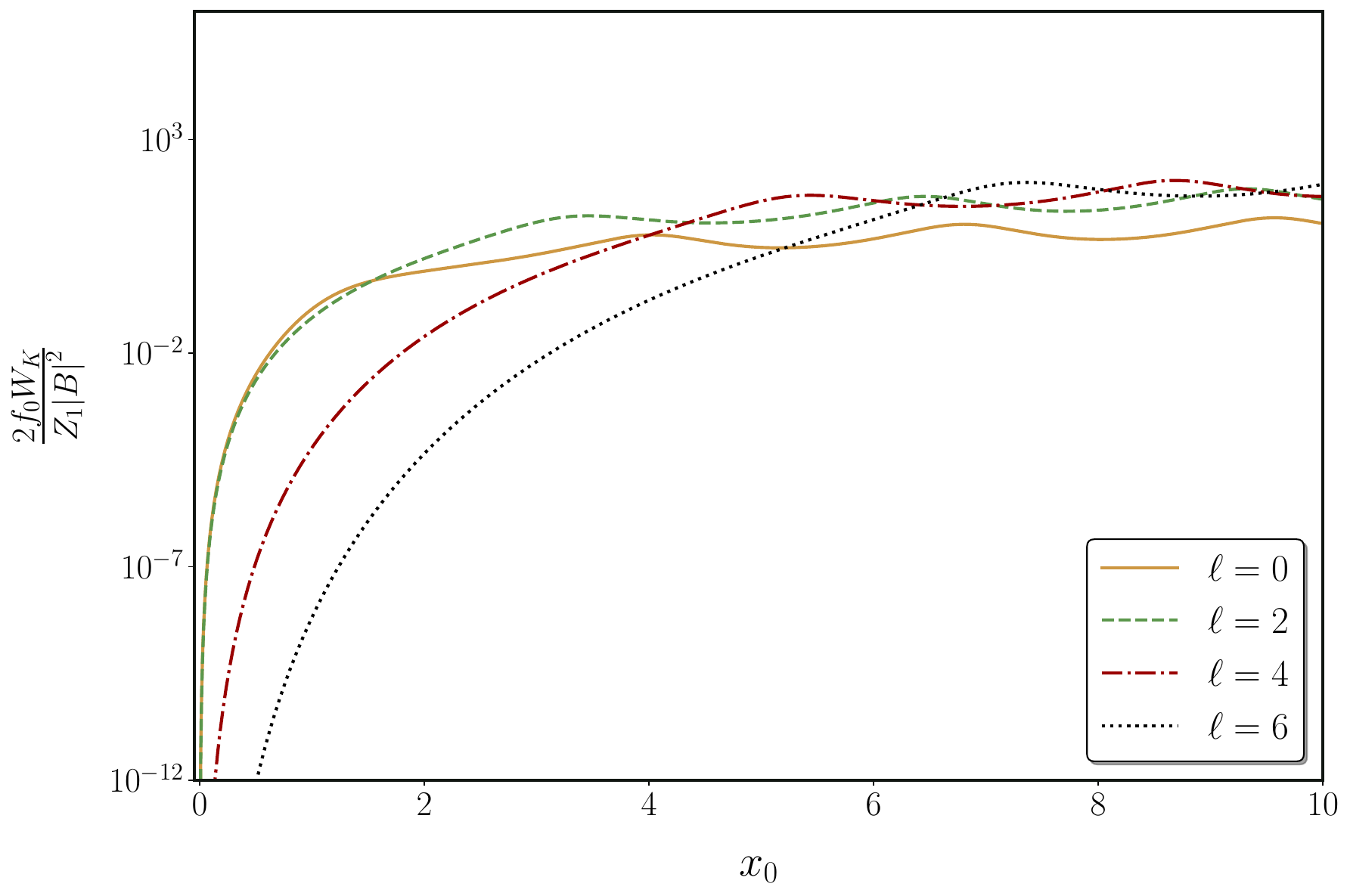}
        \includegraphics[width=0.45\textwidth]{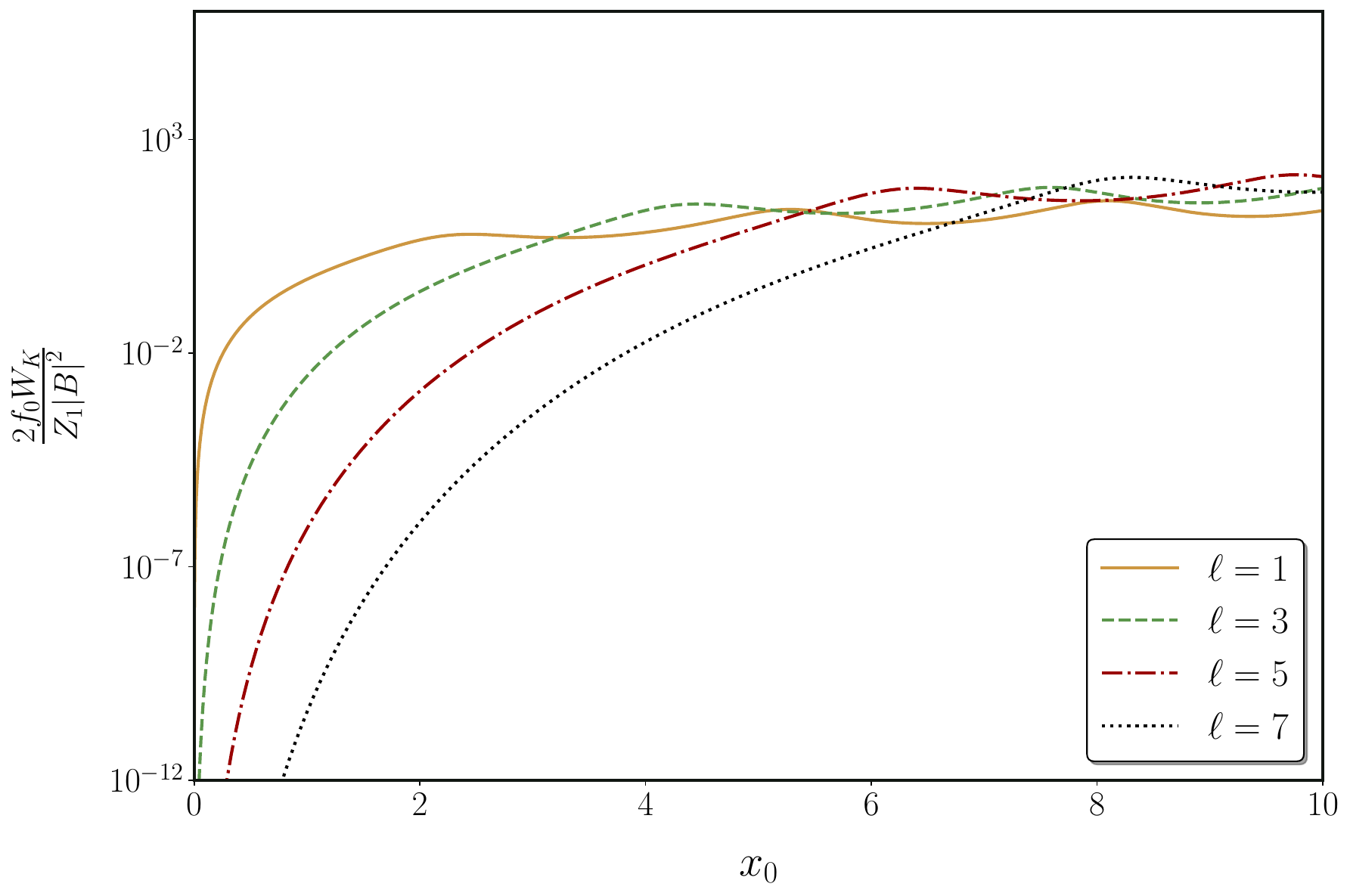}
    \caption{  \label{fig:wk2} The kinetic energy as a function of $x_0$ for the petrol bubble in seawater shows no significant resonance peaks, similar to the potential energy.}
\end{figure}

Fig.~\ref{fig:wp2} illustrates the potential energy for both even and odd values of $\ell$.
Subfigure (a) displays the results for even $\ell$ values $(0, 2, 4, 6)$, while subfigure (b) represents the odd values $(1, 3, 5, 7)$. As $x_0$ increases, the curves for all $\ell$ values converge toward a stable potential energy. 
Notably, no significant resonance peaks are observed in either the even or odd modes, indicating a lack of strong energy trapping under these specific conditions. 
This absence of resonance suggests that the moderate impedance mismatch does not induce substantial energy localization within the scattering center.

Fig.~\ref{fig:wk2} presents the kinetic energy for the same range of $x_0$ values, distinguishing between even and odd $\ell$ values. 
The trends mirror those observed in the potential energy curves, with kinetic energy remaining relatively stable, lacking pronounced oscillations or resonance peaks. 
Although minor fluctuations occur, the curves ultimately converge as $x_0$ increases.

For both potential and kinetic energy (see Figure~\ref{fig:wpwk1-1} bottom), the absence of significant resonance peaks emphasizes the role of the impedance ratio in controlling energy distribution. 
The lack of pronounced energy localization for either even or odd $\ell$ values supports the conclusion that moderate impedance differences between media prevent strong resonance effects, which are typically observed when there is a larger contrast in acoustic properties.

\section{Conclusion}
\label{sec:conclusions}

In this study, we present an analytical approach to the acoustic scattering by spherical objects, based on internal energy analysis. The motivations comes from experiments on isothermal viral inactivation by ultrasound (5--10 MHz), driven primarily by structural resonances in absence of bubble cavitation or related effects. Our findings are consistent with this hypothesis, demonstrating that resonances appear for materials with contrasting acoustic impedance. For materials with similar densities, resonances are supported only if the sound speed of the scattering center is much smaller than that of the surrounding fluid $c_1 \ll c_0$, implying a softer medium with reduced bulk modulus. Advances in acoustic metamaterials have shown that folded materials exhibit reduced sound speeds, owning to their less rigid, spring-like structure, in agreement with the organization of viral nucleoproteins.

Our numerical findings also suggest the emergence of dynamical dissipation mechanisms, even in absence of viscosity, driven by the internal localization of compressional waves. Analogous phenomena are well-documented in optics: once trapped, the waves escape after a characteristic delay $\tau_{\textrm{Wigner}}$, broadening and limiting resonances. As a result, the internal pressure field near resonances receives significant contributions from various modes, with the dominant one characterizing the center of the peak in the small particle regime. This behavior contrasts with the usual monopole-only approximation, which accounts only for radial deformations. Higher-order modes introduce heterogeneous deformations and associated shear forces. In the diffraction regime, even (odd) modes contribute nearly equally, producing a comb-like pattern, confirming that $\ell > 0$ must be taken into account. Taken together, monopole resonances in the 1–-10 MHz range are relevant for viral particles of 100–-200 nm, while higher-order modes dominate in the 10–-100 MHz range, indicating two distinct operating regimes. The lower-frequency resonances are fewer and more selective, with their positions strongly dependent on the material properties of the target particle. At higher frequencies, the density of resonances increases, leading to overlapping modes that enhance energy deposition with focus on geometrical aspects.

Our model disregards viscosity as a first approximation, although it is a key aspect in low Reynold flows such as biological systems. Introducing viscosity $\eta$ affects our model in several ways. First, the wave number $k$ becomes a complex to capture the energy loss and a frequency-dependent sound speed $c=c(k)$. Second, the scattering center behaves as a viscoelastic material, capable of both storing and dissipating energy. The internal stress $\sigma_1$ acquires a non-trivial spatial distribution, strongly dependent on the velocity field $\mathbf{v}_1(\mathbf{r},t)$. In general, the stress decays according to the relaxation time $\tau_{ \textrm{relax}} = \eta/ E$ where $E$ is the Young modulus. When the stress exceeds the molecular forces that bind them structure together, the medium breaks apart. 

The framework developed here provides a foundation for further investigations into more complex scattering geometries, such as core-shell structures, where interactions between materials could enhance energy localization. Another approach is to explore multiple scatterers, non-ideal fluid dynamics, or non-linear effects. These developments would expand the model's applicability and deepen the understanding of wave scattering across diverse physical contexts. Furthermore, the inclusion of dissipation to our current model would be beneficial for applications in biological settings, where non-inertial forces and stress propagation play crucial roles.

\begin{acknowledgments}
The authors gratefully acknowledge the financial support provided by the São Paulo Research Foundation (FAPESP) and the National Council for Scientific and Technological Development (CNPq). NER was supported by CNPq grant number 140549/2022-6; 
GN was supported by FAPESP grant 2023/07241-5; OMB thanks FAPESP by the grants 2018/22214-6, 2021/08325-2 and CNPq by the grant 307897/2018-4. ZRA  was supported by FAPESP grants 	17/09354-0 and 18/21694-4; ASM acknowledges the CNPq grant 0304972/2022-3. This support was essential for the successful completion of this research.
\end{acknowledgments}

%

\end{document}